\documentclass[
article,
superscriptaddress,
amsmath,amssymb,
aps,
]{revtex4-2}

\usepackage[utf8]{inputenc}
\usepackage{mathtools}
\usepackage{amsmath}
\usepackage{inputenc}
\usepackage{graphicx,subfigure}
\usepackage{dcolumn}%
\usepackage{bm}

\begin{document}
	
	\preprint{APS/123-QED}
	
	\title{Estimating energy levels  of a three-level atom in single and multi-parameter metrological schemes}

	\author{Hossein Rangani Jahromi}
	\email{h.ranganijahromi@jahromu.ac.ir}
	\affiliation{
		Physics Department, Faculty of Sciences, Jahrom University, P.B. 74135111, Jahrom, Iran}

	\author{Roya Radgohar}
		\affiliation{
		Physics Department, Faculty of Sciences, Jahrom University, P.B. 74135111, Jahrom, Iran}
	\affiliation{Department of Physics, Sharif University of Technology, Tehran 14588-89694, Iran
	}

	\author{Seyed Mohammad Hosseiny}
	
	\affiliation{Physics Department, Faculty of Sciences, Urmia University,P.B. 165, Urmia, Iran}

	\author{Mahdi Amniat-Talab}
	\affiliation{Physics Department, Faculty of Sciences, Urmia University,P.B. 165, Urmia, Iran}

	\date{\today}
	
	\begin{abstract}
		Determining the energy levels  of a quantum system is a significant task, for instance, to analyze reaction
		rates in drug discovery and catalysis or characterize the compatibility of materials. In this paper we exploit
		quantum metrology, the research field focusing on the estimation of
		unknown parameters exploiting quantum resources, to address this problem for a three-level system interacting with laser fields.  The performance of simultaneous estimation of the levels compared to independent one is also investigated in various scenarios. Moreover, we introduce, the Hilbert-Schmidt speed (HSS), a special type of
		quantum statistical speed, as a powerful figure of merit for enhancing estimation of energy spectrum. This measure is easily computable, 
		because it does not require diagonalization of the system state, verifying its efficiency in high-dimensional systems.
	
	\end{abstract}

	\maketitle

	\section{\label{sec:level1}Introduction}
	One of  the most important applications of physical science is the measurement
	process whose goal is to associate a value with a physical
	quantity, resulting in an estimate of it. Together with each experimental
	estimate, an uncertainty, affecting the measurement result, appears.   This statistical
	error can have two different
	natures: fundamental and technical. There are fundamental limits
	on uncertainty, such as those due to Heisenberg relations, that are
	imposed by physical laws. Conversely, the technical one is mostly represented by the accidental error, because of out-of-control imperfections in
	the measurement process.
	Since quantum mechanics is the most fundamental, predictive, and successful theory describing small scale phenomena, an investigation of the measurement process as well as
	the ultimate achievable precision bounds has to be done under the
	light of such theory \cite{helstrom1976quantum,holevo2011probabilistic,lu2021incorporating,fiderer2021general,rangani2019weak,farajollahi2018estimation,jahromi2020quantum,jafarzadeh2020effects}. In fact, on the one hand, quantum theory
	determines fundamental limits on the estimate precision. On the other
	hand, for achieving such limits, quantum resources have to be
	used.

	\par Interestingly, employing quantum systems to estimate
	unknown parameters overcomes the precision limits that can be, in
	principle, achieved by applying only classical resources. This idea is at the
heart of the continuously growing research area of quantum metrology
	aiming at reaching the ultimate fundamental bounds on estimation
	precision by using quantum probes \cite{braunstein1992quantum,PhysRevLett.72.3439,lee2002quantum,giovannetti2004quantum,giovannetti2006quantum,giovannetti2011advances,paris2009quantum,jahromi2015geometric,jahromi2018parameter,polino2020photonic,sidhu_kok_2020,gianani2020assessing,jahromi2021remote}. This field of research has attracted a great deal of interest   in the last few years, leading to notable developments
	both theoretically and experimentally as reported in previous review
	papers concerning  
	quantum phase estimation problems \cite{toth2014quantum}  optical metrology \cite{demkowicz2015quantum,moreau2019imaging,xu2019sensing,pirandola2018advances,genovese2016real}, multiparameter estimation scenario \cite{albarelli2020perspective,demkowicz2020multi,szczykulska2016multi,jahromi2019multiparameter}, and metrological tasks performed
	by various physical systems \cite{pezze2018quantum,degen2017quantum,schirhagl2014nitrogen}.

	\par

	Designing efficient quantum probes to achieve enhanced sensitivity
has been recently  generalized to a multi-parameter setting in which a set of spatially
	distributed quantum sensors is applied to simultaneously estimate a number
	of distinct parameters which cannot  be measured directly or a function thereof \cite{proctor2018multiparameter,qian2019heisenberg,sekatski2020optimal,guo2020distributed,xia2020demonstration}, a setting which is
	relevant for a wide range of applications, including multi-dimensional field and gradient estimation \cite{baumgratz2016quantum,apellaniz2018precision}, nanoscale
	NMR \cite{devience2015nanoscale}, and
	quantum networks for atomic clocks \cite{komar2014quantum} or astronomical imaging \cite{khabiboulline2019optical}. Here we investigate this strategy to enhance estimation of energy levels for a three-level system in which the cancellation of absorption may occur, impacting on important concepts and phenomena \cite{scully1997quantum} such as lasing without inversion, enhancement of the index of refraction, and electromagnetically induced transparency.
	\par
Estimating and computing   energy spectrum of a Hamiltonian are  key problems in
quantum mechanics. Many physical properties of a quantum system are primarily determined  by its energy spectrum \cite{lin1993exact,jones2019variational}.  For instance, the energy
spectra of molecules give us important information on their dynamics and therefore an
understanding of such spectra is necessary for molecular design \cite{de2016role}.
Although the ground state problem has received considerable
attention, comparing its theoretical and experimental findings with ones developed for  finding the excited states of atomic and molecular systems, we see that the latter  has experienced  less development.
This is despite the fact that it plays a key role in analyzing chemical
reactions,  a vital ingredient in the quest to discover
new drugs and industrial catalysts \cite{reiher2017elucidating,jones2019variational}. 

\par
In this paper we estimate the energy spectrum of a Hamiltonian attributed to a three-level atom interacting with laser fields. We investigate the best strategies to extract the information about the energy levels in single and multi-parameter metrological schemes. Moreover, it is illustrated that the Hilbert-Schmidt speed, a quantifier of quantum statistical speed, can be used as a powerful tool to detect the atomic energy spectrum.

\par
\par This paper is organized as follows: In Sec. \ref{sec:level2}, we
 briefly review the theory of  quantum metrology,   The model is  introduced in Sec. \ref{Model}. The independent estimation of the energy levels and application of Hilbert-Schmidt speed in this scenario are discussed in   Sec. \ref{single}.  Moreover, in Sec. \ref{multi} the simultaneous estimation of energy levels is investigated.
Finally in Sec. \ref{cunclusion}, the main results are summarized.
\section{\label{sec:level2} Estimation theory}

Assuming that $ \Phi_{\lambda} $  denotes a quantum channel depending on a set of parameters $ \lambda = (\lambda_{1}, . . . ,\lambda_{n}) $, we intend to estimate them through a
 quantum probe interacting with the channel. If the probe is  prepared in the 
 initial state 
   $ \rho $, the   output $ \rho_{\lambda}=\Phi_{\lambda}(\rho) $ should be measured
   to infer an estimation of the parameter.
The measurement  correspond
to a positive operator valued measure (POVM),  a set
of positive operators  $\{\varPi_{x}\} $ satisfying the relations  $ \sum\limits_{x}^{}\varPi_{x}\varPi_{x}^{\dagger}=\mathcal{I} $ and
  $ p(x|\lambda)\equiv \prod\limits_{i}^{} p(x_{i}|\lambda)= \text{Tr}[\varPi_{x}\rho_{\lambda}]  $, representing the probability distribution
for  results $ x=(x_{1},...,x_{M}) $ of the measurements implemented $ M $-times.
In addition, according to achieved results of the measurement, the parameters can be estimated
by using an estimator   $\tilde{\lambda}(x)=(\tilde{\lambda} _{1}(x),,...,\tilde{\lambda} _{n}(x)) $.  When $ E(\tilde{\lambda})\equiv\sum\limits_{x}^{}p(x|\lambda)\tilde{\lambda}(x)=\lambda $ [i.e., its expected value
coincides with the true value of the parameter(s)], we call it  
 an unbiased estimator \cite{brida2011optimal,ciampini2016quantum}. Physically,
 the results $ x $ are  obtained by measuring a related observable $ X $ whose eigenvalues  constitute a discrete or continuous spectrum. If the eigenvalue spectrum of that observable is continuous, the summation in the above equation must be replaced by an integral.

\par 
The efficiency of an estimator can be quantified by the covariance
matrix, $ \text{Cov}[\tilde{\lambda}] $, capturing both the variance of—and
therefore the error on—each of the individual parameters,
as well as the covariance—and hence some indication of
the correlation—between them \cite{yousefjani2017estimating}.
The Cram\'{e}r-Rao
bound states that, for all unbiased estimators $ \tilde{\lambda} $, the 
covariance matrix whose elements are defined as $\text{Cov}[\tilde{\lambda}]_{ij}=E[\tilde{\lambda}_{i}\tilde{\lambda}_{j}]-E[\tilde{\lambda}_{i}]E[\tilde{\lambda}_{j}] $
satisfies the following inequality \cite{humphreys2013quantum,liu2019quantum}
\begin{equation}\label{CRAMERRAO}
	\text{Cov}[\tilde{\lambda}]\geq \dfrac{1}{M}\textbf{I}(\lambda)^{-1},
\end{equation}
in which $ M $ denotes the number of experimental runs. Moreover,  $ \textbf{I}^{-1} $ represents the inverse of the classical Fisher information (FI) matrix whose elements are defined as follows \cite{paris2009quantum,ciampini2016quantum}
\begin{equation}\label{CFIM}
	\text{I}_{ij}=\sum\limits_{x}^{}\dfrac{1}{p(x|\lambda)}\dfrac{\partial p(x|\lambda)}{\partial \lambda_{i}} \dfrac{\partial p(x|\lambda)}{\partial \lambda_{j}}.
\end{equation}
In order to compute the quantum Fisher information  matrix (QFIM) $ \textbf{F} $ defined as the upper bound of the classical one via the
matrix inequality $ \textbf{F}\geq \textbf{I}  $ \cite{ivanov2018quantum}, we first require to obtain the Symmetric Logarithmic Derivative (SLD) $ L_{\lambda_{j}} ~~ ( j =
1,2,..., p)$ satisfying the operator equation
\begin{equation}\label{SLD1}
	\frac{\partial \rho(\lambda)}{\partial \lambda_{j}}=\dfrac{1}{2}[L_{\lambda_{j}}\rho(\lambda)+\rho(\lambda)L_{\lambda_{j}}],
\end{equation}  
where $ \rho $ denotes the density operator of the system. In the next step, the  
elements of QFIM can be easily computed by
\begin{equation}\label{QFIM}
	F_{ij}=\frac{1}{2}\text{Tr}[\rho(L_{\lambda_{i}}L_{\lambda_{j}}+L_{\lambda_{j}}L_{\lambda_{i}})].
\end{equation}
The QFI associated with the ultimate precision
in the multi-parameter estimation strategy  provides
a lower bound on the covariance matrix, i.e.,
\begin{equation}\label{CRAMER}
	\text{Cov}[\tilde{\lambda}]\geq \dfrac{1}{M} \textbf{F}(\lambda)^{-1},
\end{equation}
which is called the Cramer-Rao bound (QCRB) \cite{paris2009quantum}.

If we aim at estimating each parameter  \textit{independently} (i.e., single-parameter estimation), the inequality reads \cite{braunstein1994statistical,brida2011optimal}
\begin{equation}\label{CRAMERsingle}
	(\delta \lambda_{j})_{i}\geq \dfrac{1}{MF_{jj}},
\end{equation}
in which $ \delta \lambda_{j}\equiv \text{Var}(\tilde{\lambda}_{j})\equiv E(\tilde{\lambda}_{j}^{2})-E(\tilde{\lambda}_{j})^{2} $ and 
$ F_{jj}\equiv F(\lambda_{j}) $, denoting the
QFI  associated with the parameter $ \lambda_{j} $, is given by

\begin{equation}\label{QFImohem}
	F_{\lambda_{j}} =\text{Tr}[\rho(\lambda) L_{\lambda_{j}}^{2}].
\end{equation}
In the independent estimation scenario, a lower bound on the total
variance of all  parameters, which should be estimated, can be computed by:
\begin{equation}\label{CRAMERtotal}
	\delta_{i} \equiv \sum\limits_{j}^{} (\delta \lambda_{j})_{i}\equiv \sum_{j} \dfrac{1}{MF_{jj}}
\end{equation}

However, if we want to estimate the parameters \textit{simultaneously}, the inequality for the variance of each  parameter is obtained as \cite{brida2011optimal}
\begin{equation}\label{CRAMERmulti}
	(\delta \lambda_{j})_{s}\geq \dfrac{[\textbf{F}(\lambda)^{-1}]_{jj}}{M},
\end{equation}
Accordingly, when there exists nonzero off-diagonal elements in the QFI matrix,  the uncertainty bounds for
simultaneous and independent  estimations  may be
different from each other. Moreover, in order to obtain  a lower bound on the total
variance of all parameters of interest, we can take the
 trace of both sides of Eq. (\ref{CRAMER}), leading to \cite{humphreys2013quantum}

\begin{equation}\label{CRAMERtotal}
	\delta_{s} \equiv \sum\limits_{j}^{} (\delta \lambda_{j})_{s}\equiv \text{Tr}\big(\text{Cov}[\tilde{\lambda}]\big) \geq \dfrac{1}{M} \text{Tr}\big(\textbf{F}(\lambda)^{-1}\big).
\end{equation} 
\par
In the  single parameter scenario, using the eigenvectors of the SLD operator as the POVM, one achieve that  $\text{I}=\text{F}$  and hence the QCRB (\ref{CRAMERsingle}) can be saturated \cite{ragy2016compatibility}. 
However, in the simultaneous estimation of parameters  the procedure  may be
more challenging  than the individual one.
Since the optimal measurement for a given
parameter can be constructed by  projectors corresponding to the
eigenbasis of the SLD, we can consequently  conclude that when
$ \forall ~ (\lambda_{j},\lambda_{k} ) \in \lambda:~[L_{\lambda_{j}},L_{\lambda_{k}}]=0  $,  one can find  a common eigenbasis for all SLDs and hence a common measurement optimal from the point of
view of extracting information on all parameters simultaneously. It should be noted tat it is only a sufficient but not a necessary condition. 
In detail, if the SLDs do not commute,
this does not necessarily indicate that it is not possible
 to simultaneously extract information on all parameters
with precision matching that of the individual scenario for
each \cite{ragy2016compatibility}.

\par
There is a weaker condition stating   that the multiparameter QCRB can be saturated provided that \cite{ragy2016compatibility,napoli2019towards}
\begin{equation}\label{commutator}
	\text{Tr}(\rho[L_i,L_k])=0.
\end{equation}
Therefore, not the commutator itself but only its expectation
value on the probe state is required to vanish. 

In order to compare the performance of  simultaneous estimation with that of individual one, we define the ratio \cite{yousefjani2017estimating}:
\begin{equation}
	R=\frac{\Delta_{\text{i}}}{\Delta_{\text{s}}},
\end{equation}
where the minimal total variances in the individual and simultaneous estimations are  denoted by, respectively,  $\Delta_{\text{i}}=\sum_{j} \dfrac{1}{MF_{jj}}$ and $\Delta_{\text{s}}=\dfrac{1}{Mp} \text{Tr}\big(\textbf{F}(\lambda)^{-1}\big)$ in which $ p $ denotes the number of parameters to be estimated. In fact, a simultaneous estimation
scheme requires fewer resources than the corresponding
independent estimation scheme by a factor of the number of
parameters to be estimated. Inserting $ p $ in the definition of $ \Delta_{\text{s}} $  is
required to account for this reduction in resources.  The measure changes in the range of $0\le R\le p$ where R$\mathrm{>}$1 indicates the better performance of multi-parameter estimation rather than the single one. Considering a single repetition of the measurement, we set $ M=1 $ throughout this paper.
%%%%%%%%%%%%%%%%%%%%%%%%%%%%%%%%%%%%%%%%%%%%%%%%%%%%%

	\section{Model}\label{Model}
\begin{figure}
		\centering
		\includegraphics[width=0.3\linewidth]{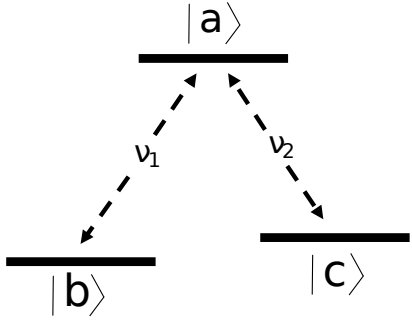}
		\caption{Three-level atom  in the $\Lambda$ configuration subject to two fields of frequencies $\nu_{1}$ and $\nu_{2}$.}
		\label{figmodel}
	\end{figure}

	We consider a  three-level atom interacting with two electromagnetic fields of frequencies ${\nu}_{1}$ and ${\nu}_{2}$ (see Fig. \ref{figmodel}). It is assumed  the atom is in  the ${\Lambda}$ configuration in which two lower levels $\left.\mathrm{\textrm{$|$}}b\right\rangle $ and$\left.\mathrm{\textrm{$|$}}c\right\rangle $ are coupled to a single upper level$\left.\mathrm{\textrm{$|$}}a\right\rangle $.
	In the rotating-wave approximation, the Hamiltonian of the system is given by \cite{scully1997quantum}:
	\begin{equation}
		\mathcal{H}={\mathcal{H}}_0+{\mathcal{H}}_1
	\end{equation}
	where 
	\begin{equation}
		\left.{\mathcal{H}}_0\mathrm{=}\mathrm{\hbar }{\mathrm{\textrm{$\omega$}}}_a\mathrm{\textrm{$|$}}a\right\rangle \left\langle a\mathrm{\textrm{$|$}}\mathrm{+}\left.\mathrm{\hbar }{\mathrm{\textrm{$\omega$}}}_b\mathrm{\textrm{$|$}}b\right\rangle \left\langle b\mathrm{\textrm{$|$}}\left.\mathrm{+}\mathrm{\hbar }{\mathrm{\textrm{$\omega$}}}_c\mathrm{\textrm{$|$}}c\right\rangle \left\langle c\mathrm{\textrm{$|$}}\right.\right.\right. 
	\end{equation}
	denotes the unperturbed Hamiltonian having eigenvalues  $ \left\{ \hbar\omega_{a},\hbar\omega_{b},\hbar\omega_{c}\right\}$, 
	and
	\begin{equation}
		{\mathcal{H}}_1\mathrm{=}{\mathrm{-}\frac{\mathrm{\hbar}}{\mathrm{2}}\mathrm{(}{\mathrm{\textrm{$\Omega$}}}_{R1}e^{-i{\textrm{$\phi$}}_1}e^{-i{\textrm{$\nu$}}_1t}\left.\mathrm{\textrm{$|$}}a\right\rangle \left\langle b\mathrm{\textrm{$|$}}\mathrm{+}{\mathrm{\textrm{$\Omega$}}}_{R2}e^{-i{\textrm{$\phi$}}_2}e^{-i{\textrm{$\nu$}}_2t}\left.\mathrm{\textrm{$|$}}a\right\rangle \left\langle c\mathrm{\textrm{$|$}}\right.\mathrm{)+H.c}\right.}
	\end{equation}
	describes the interaction between the atom and fields.  Moreover,
	${\mathrm{\textrm{$\Omega$}}}_{R1}e^{-i{\textrm{$\phi$}}_1}$ and ${\mathrm{\textrm{$\Omega$}}}_{R2}e^{-i{\textrm{$\phi$}}_2}$ represent the complex Rabi frequencies associated with the coupling of the field modes of the frequencies ${\nu}_{1}$ and ${\nu}_{2\ }$ to the atomic transitions$\left.\mathrm{\textrm{$|$}}a\right\rangle $$\mathrm{\to }$$\left.\mathrm{\textrm{$|$}}b\right\rangle $ and$\left.\mathrm{\textrm{$|$}}a\right\rangle $$\mathrm{\to }$$\left.\mathrm{\textrm{$|$}}c\right\rangle $, respectively.  We consider in which only $\left.\mathrm{\textrm{$|$}}a\right\rangle $$\mathrm{\to }$$\left.\mathrm{\textrm{$|$}}b\right\rangle $ and$\left.\mathrm{\textrm{$|$}}a\right\rangle $$\mathrm{\to }$$\left.\mathrm{\textrm{$|$}}c\right\rangle $ transitions are dipole allowed.
	\par
	Preparing the system in the initial atomic state 
	\begin{equation}
		\left.\mathrm{\textrm{$|$}}\textrm{$\psi$}(0)\right\rangle \mathrm{\ }={\mathrm{cos} \left(\frac{\textrm{$\theta$}}{2}\right)\left.\mathrm{\textrm{$|$}}b\right\rangle \ }+{\mathrm{sin} \left(\frac{\textrm{$\theta$}}{2}\right)\left.e^{-i\textrm{$\psi$}}\mathrm{\textrm{$|$}}c\right\rangle \ }, 
	\end{equation}
	which is a superposition of the two lower levels $\left\{\left.\mathrm{\textrm{$|$}}b\right\rangle,\left.\mathrm{\textrm{$|$}}c\right\rangle\right\} $, and assuming that  the fields to be resonant with the $\left.\mathrm{\textrm{$|$}}a\right\rangle $$\mathrm{\to }$$\left.\mathrm{\textrm{$|$}}b\right\rangle $ and the$\left.\mathrm{\textrm{$|$}}a\right\rangle $$\mathrm{\to }$$\left.\mathrm{\textrm{$|$}}c\right\rangle $ transitions respectively, i.e, ${\textrm{$\omega$}}_{ab}\mathrm{=}{\textrm{$\nu$}}_1$ and ${\textrm{$\omega$}}_{ac}\mathrm{=}{\textrm{$\nu$}}_2$, we find that the evolved state of the system is given by \cite{scully1997quantum,liu2019quantum}
	\begin{equation}
		\left.\mathrm{\textrm{$|$}}\textrm{$\psi$}(t)\right\rangle =\ \left.c_a\left(\mathrm{t}\right)e^{\mathrm{-}\mathrm{i}{\mathrm{\textrm{$\omega$}}}_a\mathrm{t}}\mathrm{\textrm{$|$}}a\right\rangle+\left.c_b\left(\mathrm{t}\right)e^{\mathrm{-}\mathrm{i}{\mathrm{\textrm{$\omega$}}}_b\mathrm{t}}\mathrm{\textrm{$|$}}b\right\rangle +\left.c_c\left(\mathrm{t}\right)e^{\mathrm{-}\mathrm{i}{\mathrm{\textrm{$\omega$}}}_c\mathrm{t}}\mathrm{\textrm{$|$}}c\right\rangle,
	\end{equation}
	where 
	\begin{equation}
		c_a=\frac{i~\text{sin}(\mathrm{\textrm{$\Omega$}}t/2)}{2}[{\mathrm{\textrm{$\Omega$}}}_{R1}e^{-i{\textrm{$\phi$}}_1}\mathrm{cos}\mathrm{}(\frac{\textrm{$\theta$}}{2})+{\mathrm{\textrm{$\Omega$}}}_{R2}e^{-i{(\textrm{$\phi$}}_2+\textrm{$\psi$})}\mathrm{sin}\mathrm{}(\frac{\textrm{$\theta$}}{2})] 
	\end{equation}
	\begin{equation}
		c_b=\frac{1}{{\mathrm{\textrm{$\Omega$}}}^2}\{\left[{\mathrm{\textrm{$\Omega$}}}^2_{R2}+{\mathrm{\textrm{$\Omega$}}}^2_{R1}\text{cos}\left(\frac{\mathrm{\textrm{$\Omega$}}t}{2}\right)\right]{\mathrm{cos} \left(\frac{\textrm{$\theta$}}{2}\right)\ }-2{\mathrm{\textrm{$\Omega$}}}_{R1}{\mathrm{\textrm{$\Omega$}}}_{R2}e^{i{(\textrm{$\phi$}}_1-{\textrm{$\phi$}}_2-\textrm{$\psi$})}{\text{sin}}^2\left(\frac{\mathrm{\textrm{$\Omega$}}t}{4}\right)\text{sin}\left(\frac{\mathrm{\textrm{$\theta$}}}{2}\right)\}
	\end{equation}
	\begin{equation}
		c_c=\frac{1}{{\mathrm{\textrm{$\Omega$}}}^2}\{-2{\mathrm{\textrm{$\Omega$}}}_{R1}{\mathrm{\textrm{$\Omega$}}}_{R2}e^{-i{(\textrm{$\phi$}}_1-{\textrm{$\phi$}}_2)}{\text{sin}}^2\left(\frac{\mathrm{\textrm{$\Omega$}}t}{4}\right)\text{cos}\left(\frac{\mathrm{\textrm{$\theta$}}}{2}\right)+\left[{\mathrm{\textrm{$\Omega$}}}^2_{R1}+{\mathrm{\textrm{$\Omega$}}}^2_{R2}\text{cos}\left(\frac{\mathrm{\textrm{$\Omega$}}t}{2}\right)\right]{e^{-i\textrm{$\psi$}}\mathrm{sin} \left(\frac{\textrm{$\theta$}}{2}\right)\ }\} 
	\end{equation}
	in which $\mathrm{\textrm{$\Omega$}}\mathrm{=}{{\mathrm{(}\mathrm{\textrm{$\Omega$}}}^2_{R1}+{\mathrm{\textrm{$\Omega$}}}^2_{R2})}^{\frac{1}{2}}$. In the special case that 
	\begin{equation}
		\Omega_{\mathrm{R1}}=\Omega_{\mathrm{R2}},~~\theta=\pi/2,~~\alpha=\phi_{1}-\phi_{2}-\psi=\pm \pi,
	\end{equation}
	the \textit{coherent trapping}  occurs, because
	\begin{equation}
		c_{a}(t)=0,~~c_{b}(t)=\frac{1}{\sqrt{2}},~~c_{c}(t)=\frac{1}{\sqrt{2}}\text{e}^{-i\psi},
	\end{equation}
	meaning that the population is \textit{trapped } in the lower states and there is no absorption even in the presence of the applied fields. This phenomenon arises from the quantum
	interference between the two transitions from the two lowest state to the uppermost state. Coherent population trapping (CPT)  can easily be observed in the D lines of alkali
	atoms (Cs, Rb, K, etc.) \cite{vanier2005atomic,shah2010advances}.  CPT resonance signals have been
	widely studied for precision sensing with potential applications in microwave atomic clocks
	based on vapor cells and cold atoms, magnetometers, high-resolution spectroscopy, atomic interferometers,  and phase-sensitive amplification
	\cite{zhu2000theoretical,liu2017low,schwindt2004chip,liu2013ramsey,cheng2016coherent,neveu2018phase}.
	\par
	In this paper we show that CPT may play an important role in estimation of the energy levels of an atomic system. We set $ \theta=\pi/2 $ throughout this paper.

	\section{Single parameter estimation}\label{single}
	In this section, computing the QFIs, we investigate the single parameter estimation for detecting energy levels of a three level atom in two different regimes: (a) different Rabi frequencies (${\mathrm{\textrm{$\Omega$}}}_{R1}\neq $ ${\mathrm{\textrm{$\Omega$}}}_{R2}$), and  (b) equal Rabi frequencies (${\mathrm{\textrm{$\Omega$}}}_{R1}=$ ${\mathrm{\textrm{$\Omega$}}}_{R2}$). The analytical expressions of the QFIs are given
	in the Appendix. It is interesting to see that the QFIs are independent of the energy levels as well as the transition frequencies. Moreover the QFIs do not explicitly depend on   the laser phases $(\phi _{1},\phi _{2}) $ and  the initial phase $ \psi $. Instead, they rise in defining  parameter $ \alpha=\phi _{1}-\phi _{2}-\psi $ appearing in the QFIs expressions. Accordingly the effects of the laser phases on the estimation of the energy levels can be  controlled by changing  phase $ \psi $ encoded into the initial state of the system. 
	\subsection{Single parameter estimation with respect to ${\boldsymbol{\mathrm{\textrm{$\omega$}}}}_{\boldsymbol{a}}$}
\begin{figure}[ht]
		\subfigure[]{\includegraphics[width=7cm]{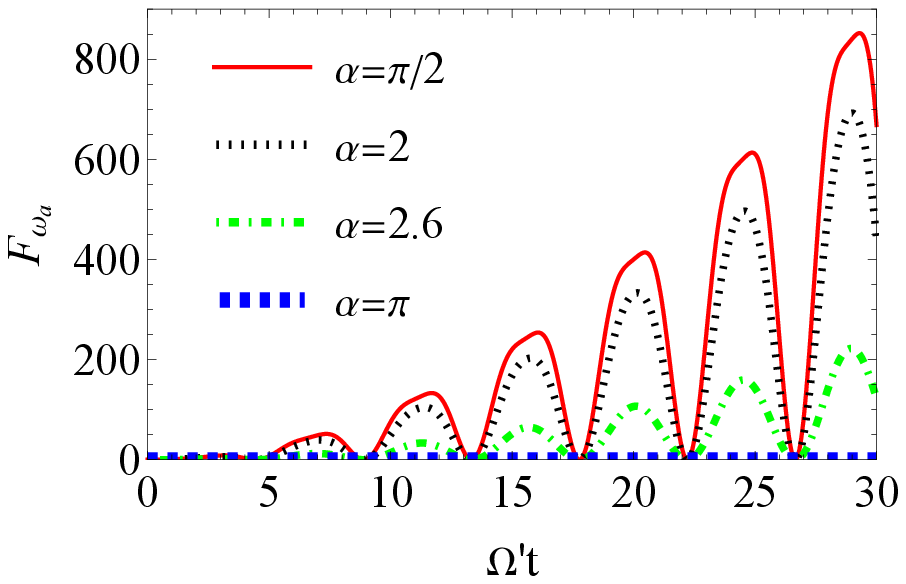}\label{fig_2a} }
		\hspace{10mm}
		\subfigure[]{\includegraphics[width=7cm]{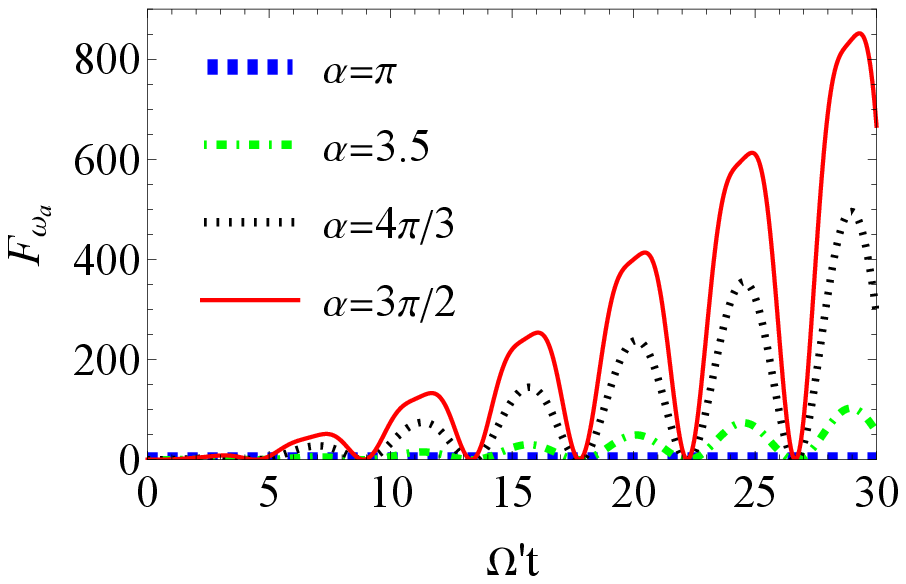}\label{fig_2b} }
		\caption{The dynamics of  ${{\mathrm{F}}}_{{{\mathrm{\textrm{$\omega$}}}}_{{\mathrm{a}}}}$ as a function of the scaled time   ${\mathrm{t}}{\mathrm{\textrm{$\Omega$}}}{\mathrm{'}}$ for ${\textrm{$\theta$}}{=}{\textrm{$\pi$}}{/}{2}$ and various values of ${\textrm{$\alpha$}}$, in equal Rabi frequencies regime ${{\mathrm{\textrm{$\Omega$}}}}_{{R}{1}}{=}{{\mathrm{\textrm{$\Omega$}}}}_{{R}{2}}{=}{\mathrm{\textrm{$\Omega$}}}{\mathrm{'}}$.\textbf{}}
		\label{fig:singleQIomegaa}
	\end{figure}

	For the case of equal Rabi frequencies $\mathrm{(}{\mathrm{\textrm{$\Omega$}}}_{R1}={\mathrm{\textrm{$\Omega$}}}_{R2}=\mathrm{\textrm{$\Omega$}}\mathrm{')}$, the dynamics of the QFI with respect to ${\mathrm{\textrm{$\omega$}}}_a$ is plotted in Fig. \ref{fig:singleQIomegaa} for different values of the parameter $ \alpha $.
	The maximized QFI, obtained for  $\textrm{$\alpha$}=\boldsymbol{\textrm{$\pi$}}\boldsymbol{/}\boldsymbol{2}$\textbf{ }and\textbf{  }$\textrm{$\theta$}=\boldsymbol{\textrm{$\pi$}}\boldsymbol{/}\boldsymbol{2}$, is given by
	
	\begin{equation}
		{F}^{\text{max}}_{{\textrm{$\omega$}}_{\mathrm{a\ }}}= \begin{array}{c}
			\frac{1}{2}t^2{\mathrm{sin}}^2(\frac{t{\mathrm{\textrm{$\Omega$}}}^\prime}{\sqrt{2}})(\mathrm{cos}(\sqrt{2}t{\mathrm{\textrm{$\Omega$}}}^\prime)+3 \end{array}).
	\end{equation}
Therefore at times $ {\mathrm{\textrm{$\Omega$}}}^\prime t=\sqrt{2}\pi $ where the QFI becomes equal to zero and hence the estimation fails. Accordingly, the Rabi frequency can be used as an efficient tool to shift the moments at which the  the single upper level is hidden from measuring devices.
\par
Moreover, 	the QFI vanishes for $ \theta=\pi/2 $ and $ \alpha=\pi $, showing that no information about the energy of the upper excited state can be extracted from the qutrit when the coherent trapping occurs. Therefore, when the population is trapped in the lower states, the information  about the single excited state becomes unaccessible. Distancing from the coherent trapping mode, for example by 
	decreasing  $ \alpha $ from $ \pi $ to $ \pi/2 $ (see Fig. \ref{fig_2a}) or increasing it in the range $ [\pi/2,3\pi/2]$ (see Fig. \ref{fig_2b}), we can enhance the parameter estimation of the  upper energy level.
	
	\subsection{  Single parameter estimation with respect to ${\boldsymbol{\mathrm{\textrm{$\omega$}}}}_{\boldsymbol{b}}$}
\begin{figure}[ht]
		\subfigure[]	{\includegraphics[width=8cm,height=5cm]{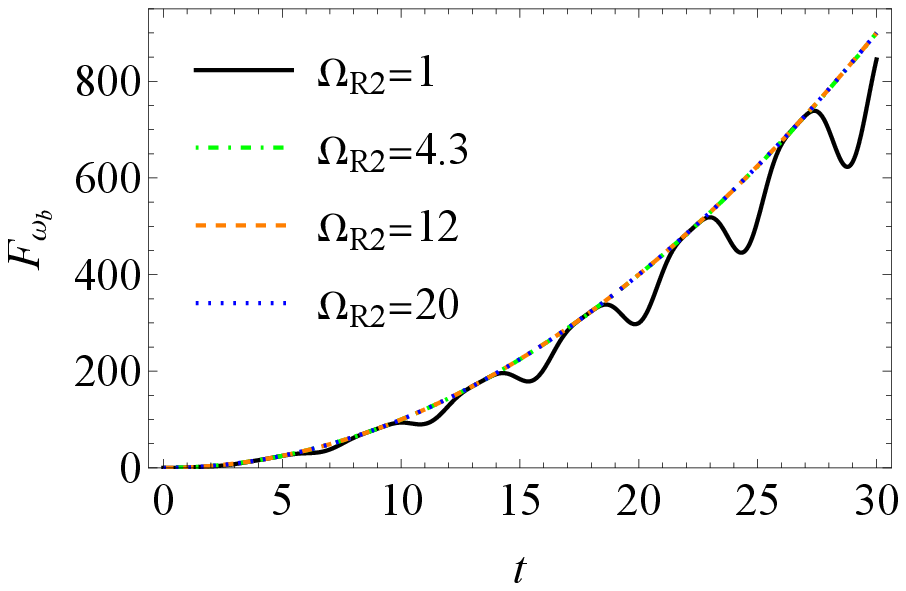}\label{fig_3a}}
		\hspace{10mm}
		\subfigure[]{	\includegraphics[width=8cm,height=5cm]{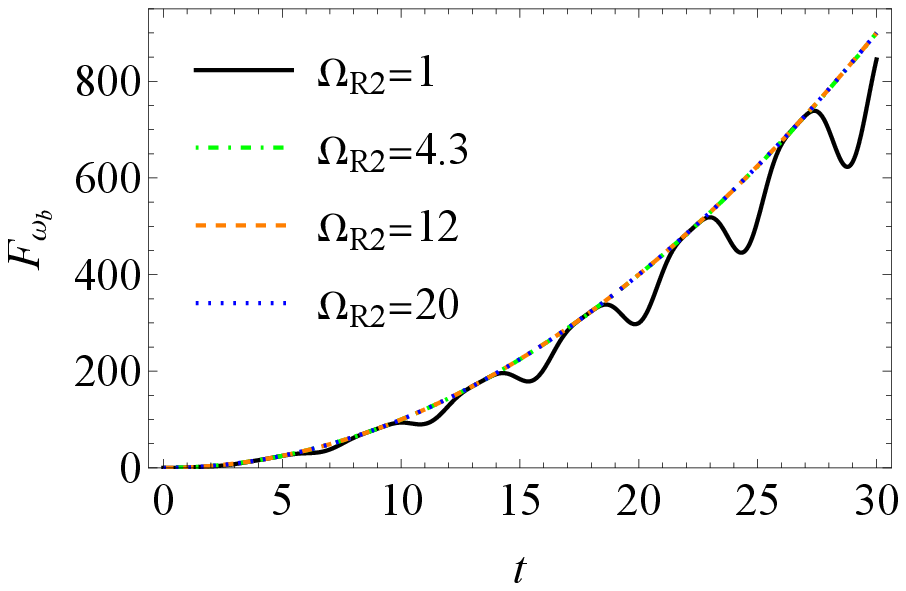}\label{fig_3b}}
		\caption{The time variation of ${{\mathrm{F}}}_{{{\mathrm{\textrm{$\omega$}}}}_{{\mathrm{b}}}}$ for ${\mathrm{\ }}{{\mathrm{\textrm{$\Omega$}}}}_{{R}{1}}{\mathrm{=}}{\mathrm{1}}$\textbf{, }${\textrm{$\alpha$}}{=}{\textrm{$\pi$}}{/}{2}$, ${\textrm{$\theta$}}{=}{\textrm{$\pi$}}{/}{2}$ and various values of ${{\mathrm{\textrm{$\Omega$}}}}_{{R}{2}}$.}
		\label{fig:QFIomegabOR}
	\end{figure}

	Another interesting scenario is quantum estimation of the lowest level energy $ \omega_{b} $. Figure \ref{fig:QFIomegabOR} exhibits the QFI dynamics in the  different Rabi frequencies regime (${\mathrm{\textrm{$\Omega$}}}_{R1}\neq $ ${\mathrm{\textrm{$\Omega$}}}_{R2}$) for various values of ${\mathrm{\textrm{$\Omega$}}}_{R2}$. As observed in 
	Figs. \ref{fig_3a} and \ref{fig_3b}, an increase 
	in the Rabi frequency ${\mathrm{\textrm{$\Omega$}}}_{R2}$  decreases the amplitude of the time oscillations of the QFI such that for large values of ${\mathrm{\textrm{$\Omega$}}}_{R2}$,  ${\mathrm{QFI}}_{{\mathrm{\textrm{$\omega$}}}_{\mathrm{a}}}$ monotonically increases with time and the best estimation occurs. In particular, we find that 
	\begin{equation}
		\lim_{\Omega_{\mathrm{R2}} \to \infty}F_{\omega_{b}}=t^{2}.
	\end{equation}
	
	\par Because the Rabi frequency  is proportional to the field amplitude, we find that intensifying the field, being resonant with the allowed transition between the intermediate  levels, one can  achieve the best precision in quantum estimation of the lowest level energy. This situation is similar to what occurs in  \textit{electromagnetically
		induced transparency} in which one of the two fields, driving the upper two levels, is  intense whereas the
	other is weak and can be treated as a probe field. Under appropriate parametric conditions, it  is possible to show that the system, independently
	of the initial condition,  evolves to a stationary state
	in which it is transparent (zero absorption) to the probe field. Therefore, the population of the lower levels increase with time, leading to a monotonous  increase in precision of estimating the ground state energy of the atom. 
	\begin{figure}[ht]
		\subfigure[] {\includegraphics[width=8cm,height=5cm]{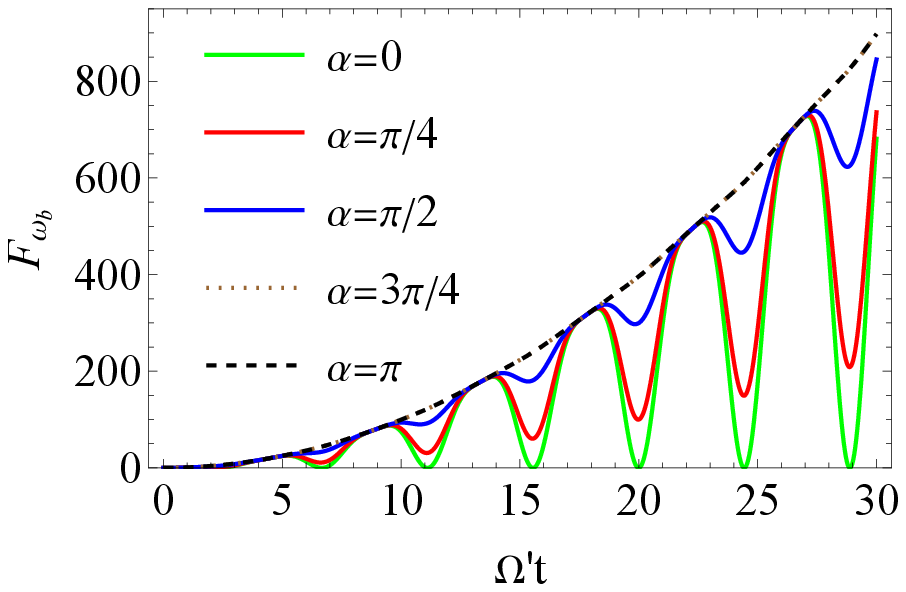}\label{fig_4a}}	
		\hspace{10mm}
		\subfigure[] {	\includegraphics[width=8cm,height=5cm]{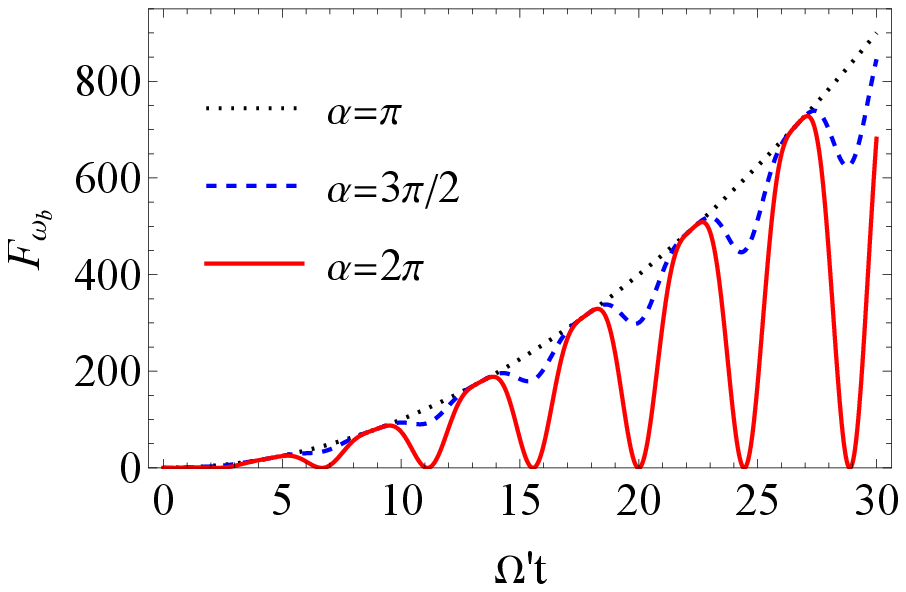}\label{fig_4b}}
		\caption{The variation of ${{\mathrm{F}}}_{{{\mathrm{\textrm{$\omega$}}}}_{{\mathrm{b}}}}$  as a function of the scaled time   ${\mathrm{t}}{\mathrm{\textrm{$\Omega$}}}{\mathrm{'}}$ for ${\textrm{$\theta$}}{=}{\textrm{$\pi$}}{/}{2}$ and  different values of ${\textrm{$\alpha$}}$  in the equal Rabi frequencies regime, i.e., ${{\mathrm{\textrm{$\Omega$}}}}_{{R}{1}}{=}{{\mathrm{\textrm{$\Omega$}}}}_{{R}{2}}{=}{\mathrm{\textrm{$\Omega$}}}{\mathrm{'}}$.}
		\label{figQFIomegab}
	\end{figure}
	
	In Figs. \ref{fig_4a} and \ref{fig_4b}, we show the time evolution of the QFI in equal Rabi frequencies $\mathrm{(}{\mathrm{\textrm{$\Omega$}}}_{R1}={\mathrm{\textrm{$\Omega$}}}_{R2}=\mathrm{\textrm{$\Omega$}}\mathrm{')}$ for different values of  $\textrm{$\alpha$}$.  We observe that 
	the best estimation is realized when the coherent trapping occurs, leading to the following expression for the optimal QFI:
	\begin{equation}
	F^{max}_{{\textrm{$\omega$}}_{\mathrm{b\ }}}=t^2,
	\end{equation}
	verifying monotonic  enhancement of parameter estimation with time.

	\subsection{  Single parameter estimation with respect to ${\boldsymbol{\mathrm{\textrm{$\omega$}}}}_{\boldsymbol{c}}$}
	Investigating the single-parameter estimation with respect to 
	$ \omega_{c} $, we find that  an increase 
	in the Rabi frequency ${\mathrm{\textrm{$\Omega$}}}_{R1}$  decreases the amplitude of the time oscillations of the QFI such that for large values of ${\mathrm{\textrm{$\Omega$}}}_{R1}$, we have 
	\begin{equation}
		\lim_{\Omega_{\mathrm{R1}} \to \infty}F_{\omega_{c}}=t^{2}.
	\end{equation}
	
	\par As discussed in the previous section the Rabi frequency  is proportional to the field amplitude, and hence we conclude that  intensifying the field, being resonant with the allowed transition between the lower ground  and  upper excited  states, we can  optimally estimate the energy of  the intermediate level. 
	
	Again, we find that  in equal Rabi frequencies regime when the coherent trapping 
	is realized, the best estimation is achieved such that the optimal QFI is given by:
	\begin{equation}
		F^{\text{max}}_{{\textrm{$\omega$}}_{\mathrm{c\ }}}=t^2,
	\end{equation}
	showing monotonic  enhancement of parameter estimation with time.

		\subsection{ Hilbert-Schmidt as an efficient tool for estimation of energy levels}
		\begin{figure}[ht]
			\includegraphics[width=5cm,height=4cm]{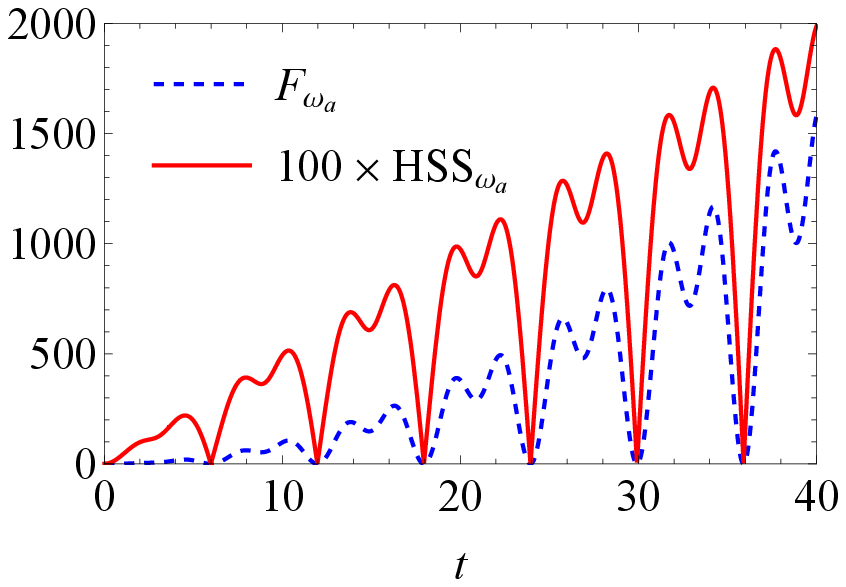}
			\hspace{8mm}
			\includegraphics[width=5cm,height=4cm]{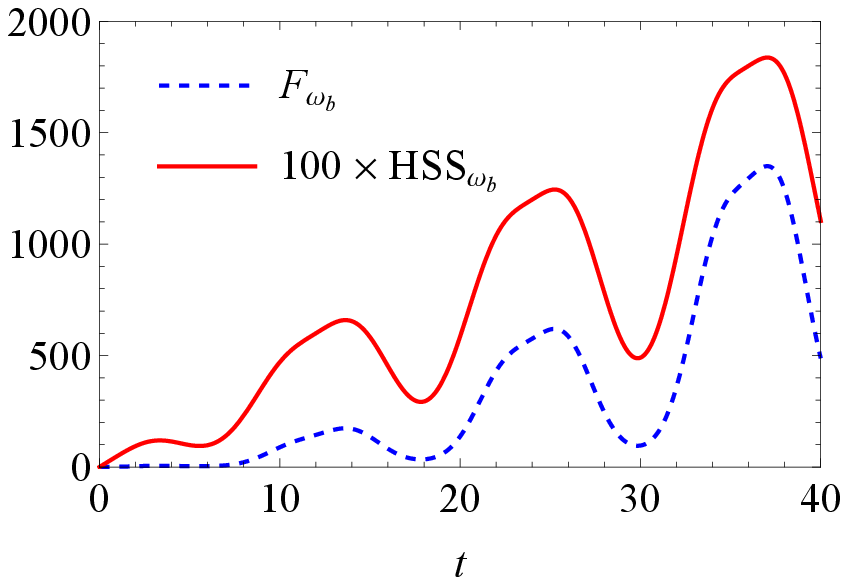}
			\hspace{8mm}
			\includegraphics[width=5cm,height=4cm]{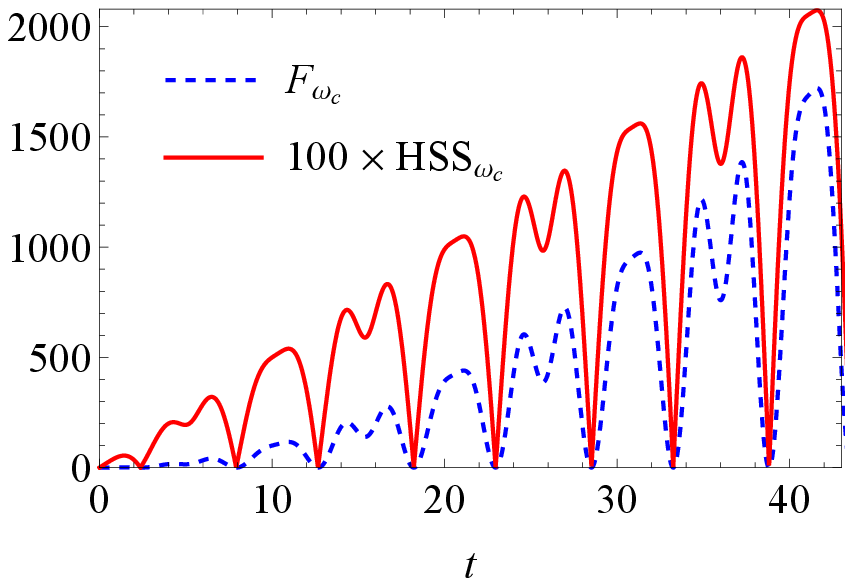}
			\caption{Comparison between dynamics of quantum Fisher information $ F_{\omega_{i}} $ and Hilbert-Schmidt speed $ HSS_{\omega_{i}} $ (amplified by 100 times)  for ${{\mathrm{\textrm{$\Omega$}}}}_{{R}{1}}{\mathrm{=}}{\mathrm{0}}.{\mathrm{32,}}$ ${{\mathrm{\textrm{$\Omega$}}}}_{{R}{2}}=1,$${\mathrm{\ }}$${{\mathrm{\textrm{$\phi$}}}}_{{1}}{\mathrm{=}}{\mathrm{1}},{{\mathrm{\textrm{$\phi$}}}}_{{2}}{\mathrm{=}}{\mathrm{1}},{\mathrm{\textrm{$\psi$}}}{=}{2}{\textrm{$\pi$}}{\ }{and}{\ }{\textrm{$\theta$}}{=}{\textrm{$\pi$}}{/}{2}.$ }
			\label{Hilbert1}
		\end{figure}
		 Given quantum state $ \rho(\varphi)$, one can define the   Hilbert–Schmidt speed (HSS) given by \cite{gessner2018statistical,jahromi2020witnessing}
		  \begin{align}\label{HSSS}
		  	HSS_{{\varphi}}(\rho\big)=\sqrt{\frac{1}{2}\text{Tr}\bigg[\bigg(\dfrac{d\rho(\varphi)}{d\varphi}\bigg)^2\bigg]},
		  \end{align} 
		   is a special quantifer of quantum statistical speed associated with  parameter $\varphi  $.
		It can be easily computed without  diagonalization of  $ \text{d}\rho(\varphi)/\text{d}\varphi $.
		\par
		In Ref. \cite{jahromi2021hilbert}, the HSS has been introduced as  a powerful figure of merit for enhancing quantum phase estimation in an open quantum system 
		made of $ n $ qubits.  More generally, because both QFI and HSS are quantum statistical speeds associated, respectively, with the \textit{Bures} and \textit{Hilbert–Schmidt distances}, it is reasonable to explore how they can be 
		related to each other. Accordingly, here we investigate the application of the HSS in detecting the energy levels of our three-level system.
\par
The dynamical comparison between the the QFI and HSS computed with respect to the same parameters, is performed numerically. In particular, we show 
in Fig. \ref{Hilbert1} that both the QFI and HSS dynamics simultaneously exhibit an oscillatory behavior such that their 
maximum and minimum points exactly coincide. These plots qualitatively verifies  that the HSS can 
detect exactly the times at which the best  estimation of the energy levels occurs (maximum of the QFI).

\par
 Because the HSS is an easily 
computable quantity having the advantage of avoiding diagonalization of the evolved density matrix, our result shows that it can be used  as a convenient  as well as powerful 
figure of merit in estimating the energy levels.

	\section{Multi-parameter estimation of energy levels}\label{multi}

As we discussed in Sec. \ref{sec:level2} if 
	$ \forall ~ (\omega_{i},\omega_{j} )  :~[L_{\omega_{i}},L_{\omega_{j}}]=0  $ where $ i,j=a,b,c $, then there is no additional
	difficulty in extracting optimal information from a state
	on all energy levels simultaneously. Although the SLDs associated with the  parameters $ \omega_{a}, \omega_{b}, \omega_{c} $ do not satisfy the above relation, 
	we fortunately find that  the commutation  condition (\ref{commutator}) holds in our model. This means  that  there is a
	single measurement which is jointly optimal for extracting information on all  energy levels, which should be estimated simultaneously, from the output state, ensuring
	the asymptotic saturability of the QCRB.
	
		\subsection{Single-qutrit scenario}
	
Employing a single-qutrit system in the estimation process, we analyze the variance of estimating one of the atomic energy levels   when it is  estimated  simultaneously. 
 In the equal Rabi frequencies regimes, simultaneously estimating $ \omega_{a} $ and $ \omega_{b} $, we find that the qualitative results extracting for estimation variance of $ \omega_{a} $ ($ \omega_{b} $), are exactly similar to ones achieved from Fig. \ref{fig:singleQIomegaa}   (\ref{figQFIomegab}) for single parameter estimation, i.e.; (a) Distancing from the coherent trapping
mode, we can enhance the parameter estimation of the upper energy level; 
(b) The best estimation of the ground state energy is realized when the coherent trapping occurs.

	\subsection{Two-qutrit scenario}

	\begin{figure}[ht]
		\includegraphics[width=8cm,height=5cm]{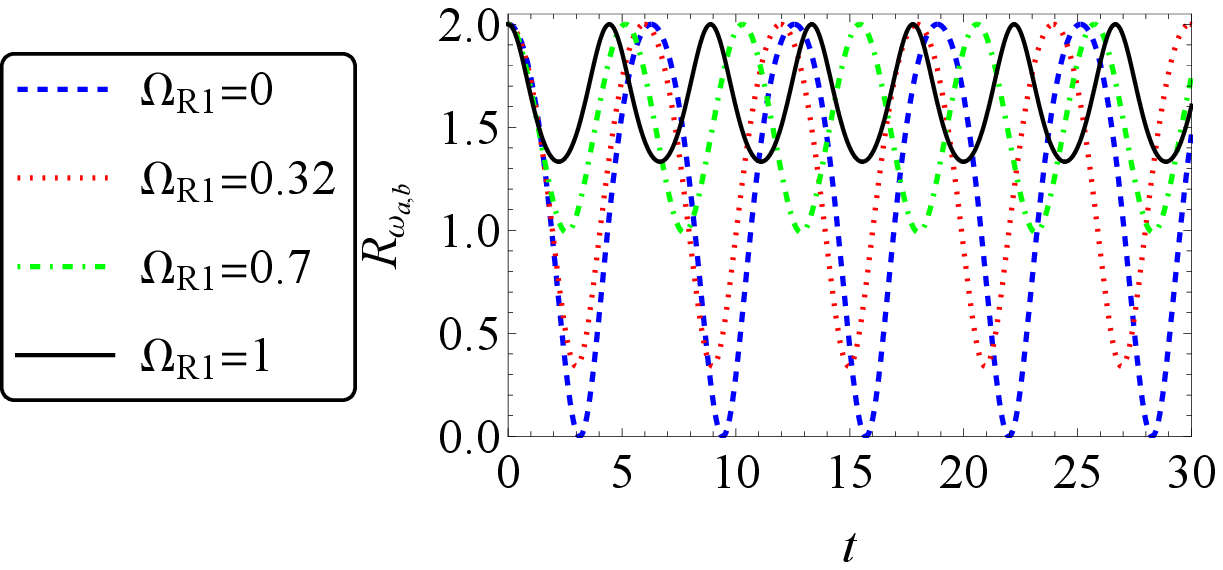}
		\hspace{10mm}
		\includegraphics[width=8cm,height=5cm]{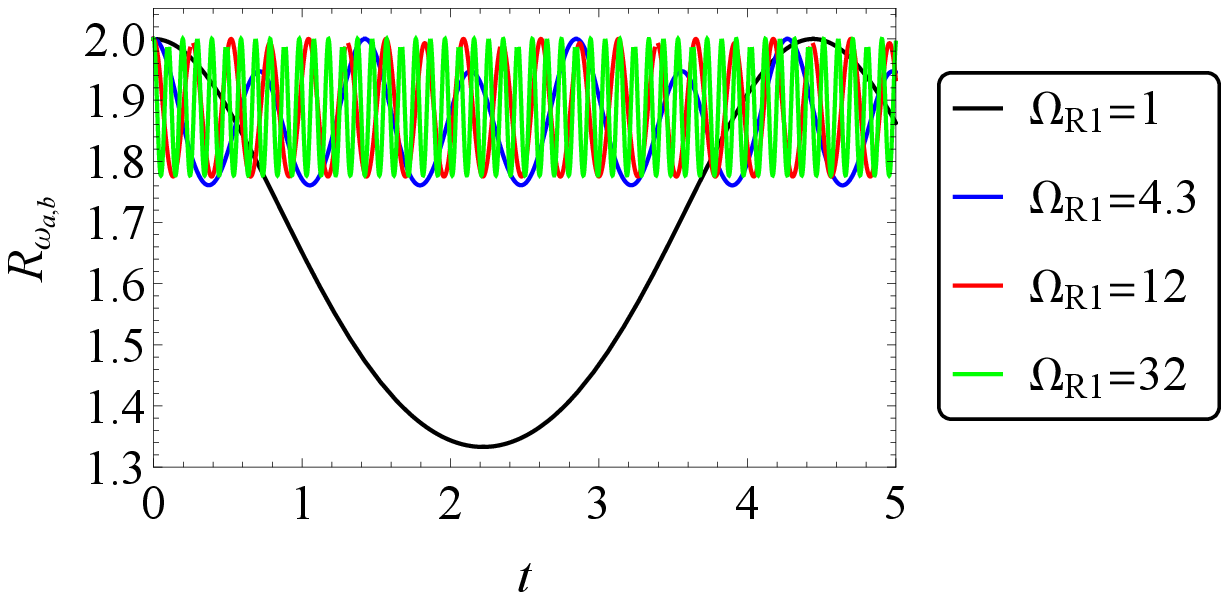}
		\caption{The dynamics  of
		performance ratio $R_{{\textrm{$\omega$}}_{\mathrm{a,b}}}$, computed for simultaneous estimation of $ \omega_{a} $ and $ \omega_{b} $,  for ${\textrm{$\alpha$}},{\textrm{$\theta$}}{=}{\textrm{$\pi$}}{/}{2}$, $ \Omega_{\mathrm{R2}}=1 $ and  various values of  ${{\mathrm{\textrm{$\Omega$}}}}_{{R}{1}}$.}
		\label{Romegaab1}
	\end{figure}

		\begin{figure}[ht]
		\includegraphics[width=8cm,height=5cm]{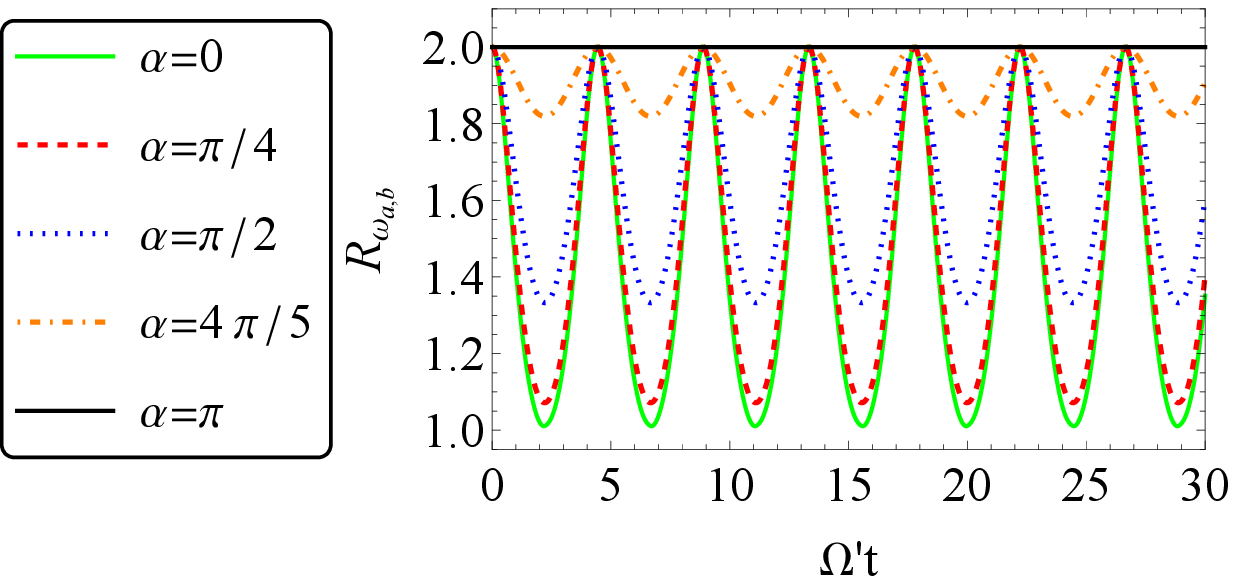}
		\hspace{10mm}
		\includegraphics[width=8cm,height=5cm]{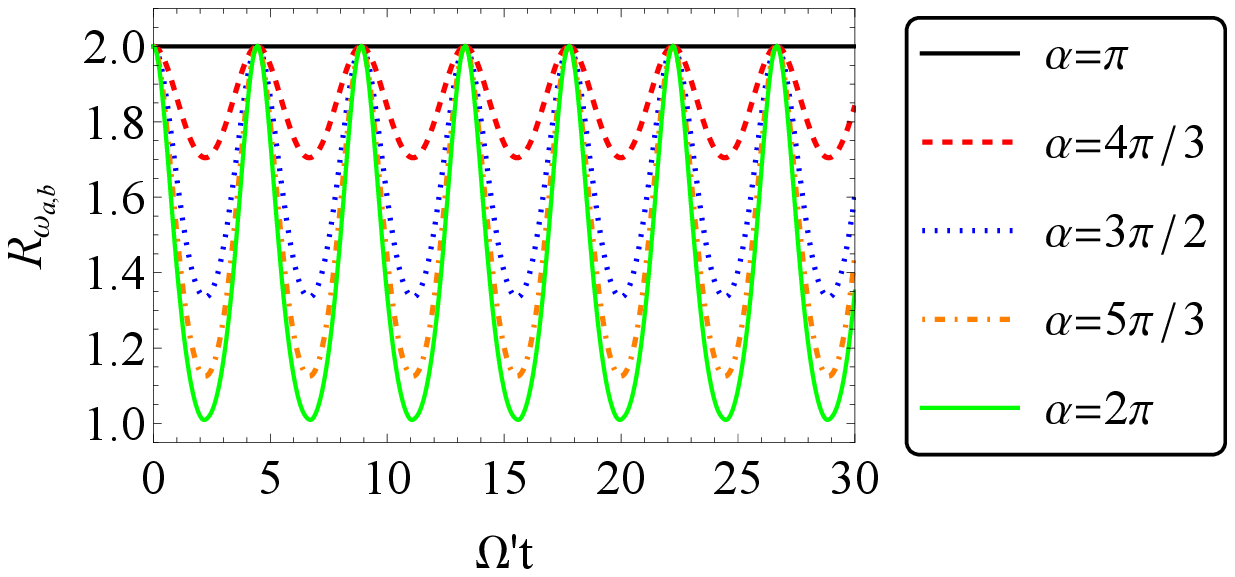}
		\caption{The variation  of
			performance ratio $R_{{\textrm{$\omega$}}_{\mathrm{a,b}}}$, calculated for simultaneous estimation of $ \omega_{a} $ and $ \omega_{b} $, as a function of the scaled time $ \Omega^{'}t $
			for ${\textrm{$\theta$}}$${=}{\textrm{$\pi$}}{/}{2}$
			and different values of
			${\textrm{$\alpha$}}$  in the equal Rabi frequencies regime.}
		\label{Rab}
	\end{figure}
	
	It is clear that when a single qutrit is applied for estimation,  no advantage  is attained by simultaneously estimating the parameters. 
	 In the second strategy we are interested to investigate what happens if we increase the number of resources. In detail, we assume that
	two uncorrelated
	qubtrits are applied in the estimation scheme.  In the individual  estimation one of which is
	used to estimate $ \omega_{i} $ and the other is used to estimate $ \omega_{j} $ where $ i,j=a,b,c $, leading to  the minimal total variance on both the parameters as
	$ \Delta_{i}(2)=(1/F_{ii})+(1/F_{jj}) $. However, in the simultaneous estimation each of these qutrits is used to estimate both
	parameters, and hence the results from the two qutrits can be classically combined
	to achieve 	$ \Delta_{s}(2)=\frac{1}{2}\Delta_{s}(1)=\frac{1}{2}([\textbf{F}(\lambda)^{-1}]_{ii}+[\textbf{F}(\lambda)^{-1}]_{jj} )=\frac{1}{2}\text{Tr}(\textbf{F}^{-1})$.

	\subsubsection{Simultaneous estimation of $ \omega_{a} $ and $ \omega_{b} $}
	
	In the first interesting scenario in which the lowest and highest energy levels, i.e., $\{\omega_{a}, \omega_{b} \} $,  should be estimated, we see that the ratio $ R_{\omega_{a,b}} $ may become greater than one  by an increase in the Rabi frequency $ \Omega_{\mathrm{R1}} $ (see Fig. \ref{Romegaab1}). Therefore,   imposing an intense field  resonant with the allowed transition between the lower ground and upper excited
	states, we observe that  the simultaneous strategy 
	always advantageous than the individual one. 
	\par
		
		If
	the off-diagonal elements of
	the QFIM are zero, the parameters of interest
	would are statistically independent, meaning that the indeterminacy
	of one of which does not affect the error on estimating the others. Under this condition
	$ \Delta_{s}(2) =\frac{1}{2}\Delta_{i}(2)$, and hence the ratio
	$ R=\Delta_{i}/\Delta_{s} $  have a maximum value
	of $ 2 $, indicating complete superiority of the simultaneous
	scheme. Figure \ref{Rab} illustrates that   in the equal Rabi frequencies regime  when the coherent trapping occurs, the QFIM becomes diagonal. Consequently, $ R_{\omega_{a,b}} $ achieved its maximum value and hence the  superiority of the simultaneous estimation is completely realized. Deviating from the coherent trapping by changing $ \alpha $, we find that the efficiency of the simultaneous estimation compared to the independent one decreases.
	\subsubsection{Simultaneous estimation of $ \omega_{a} $ and $ \omega_{c} $}
		\begin{figure}[ht]
		\includegraphics[width=8cm,height=5cm]{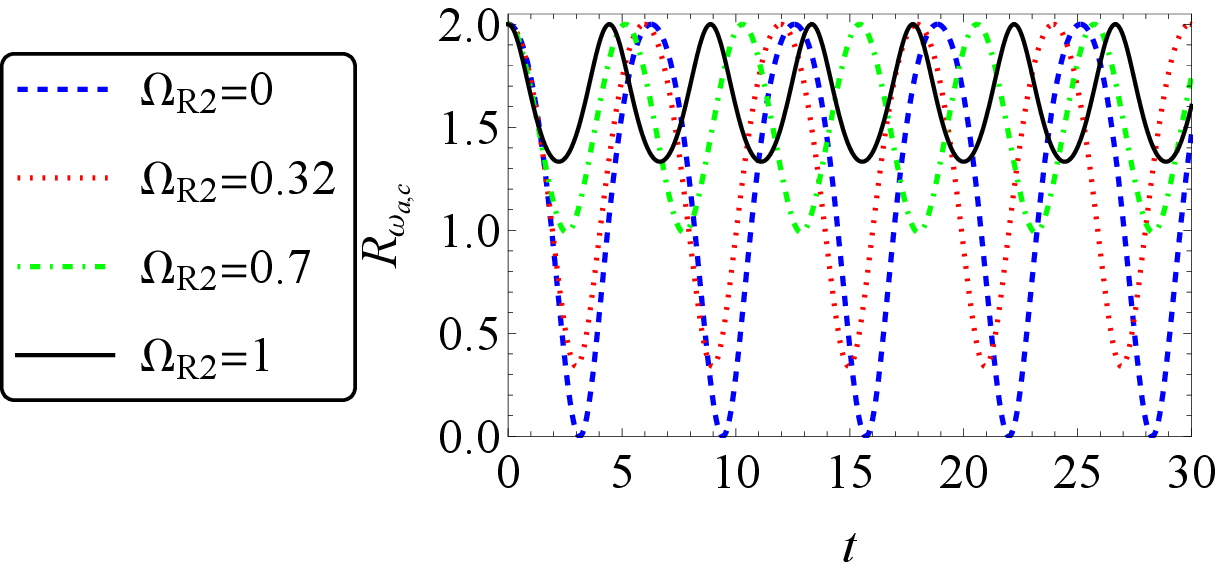}
		\hspace{10mm}
		\includegraphics[width=8cm,height=5cm]{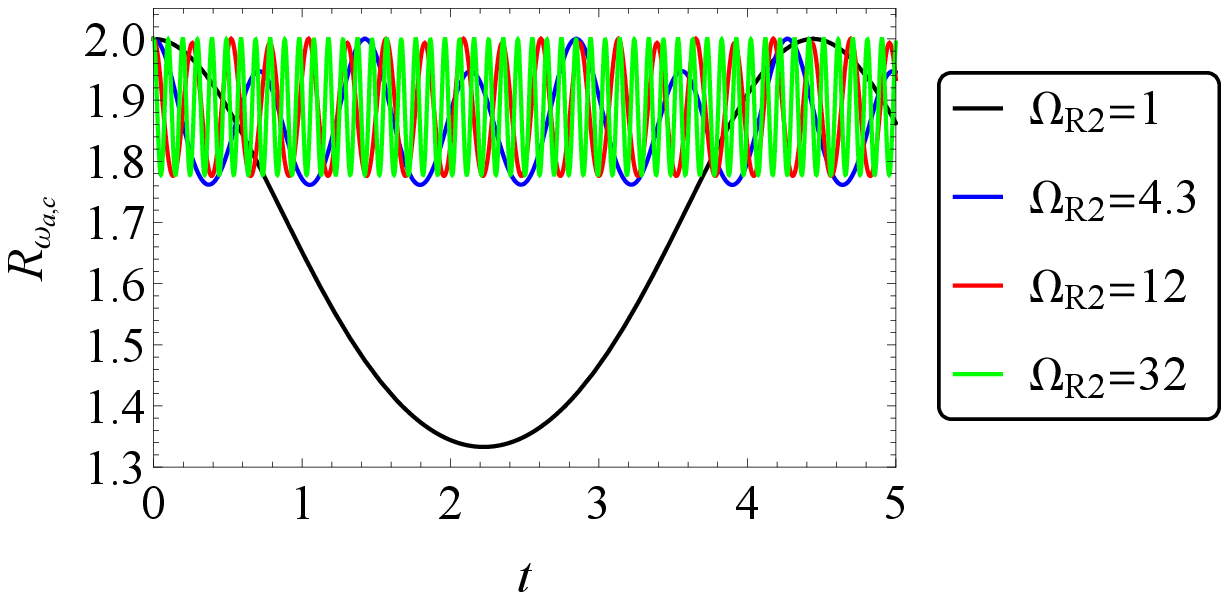}
		\caption{ The dynamics  of
			performance ratio $R_{{\textrm{$\omega$}}_{\mathrm{a,c}}}$, calculated for simultaneous estimation of $ \omega_{a} $ and $ \omega_{c} $,  for ${\textrm{$\alpha$}},{\textrm{$\theta$}}{=}{\textrm{$\pi$}}{/}{2}$, $ \Omega_{\mathrm{R1}}=1 $ and  different values of  ${{\mathrm{\textrm{$\Omega$}}}}_{{R}{2}}$.}
		\label{Romegaac1}
	\end{figure}

\begin{figure}[ht]
	\includegraphics[width=8cm,height=5cm]{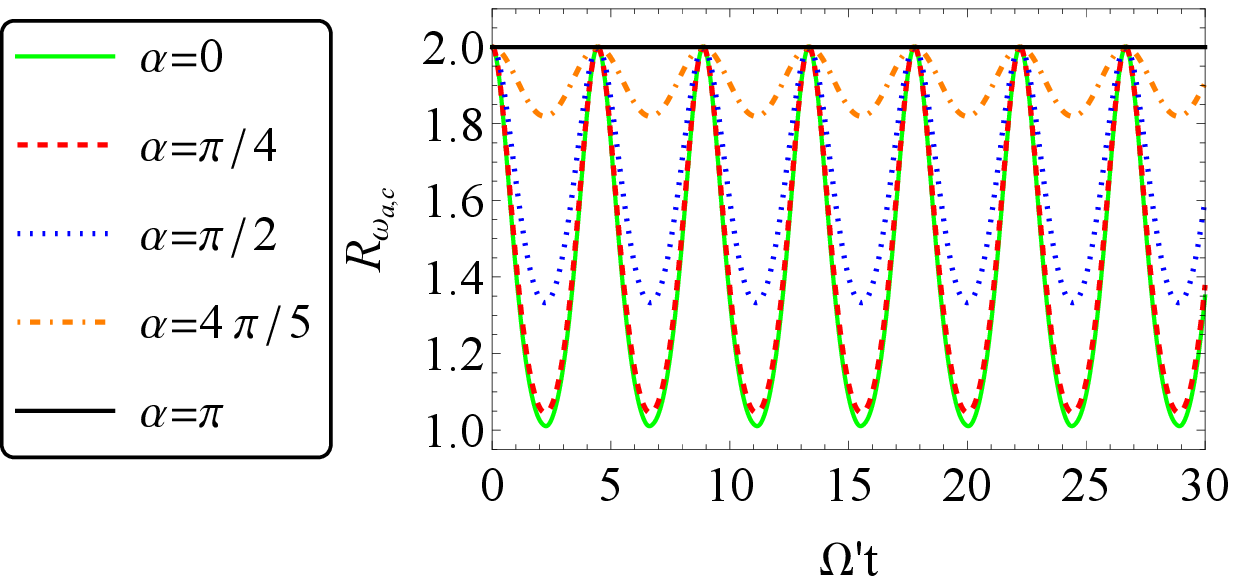}
	\hspace{10mm}
	\includegraphics[width=8cm,height=5cm]{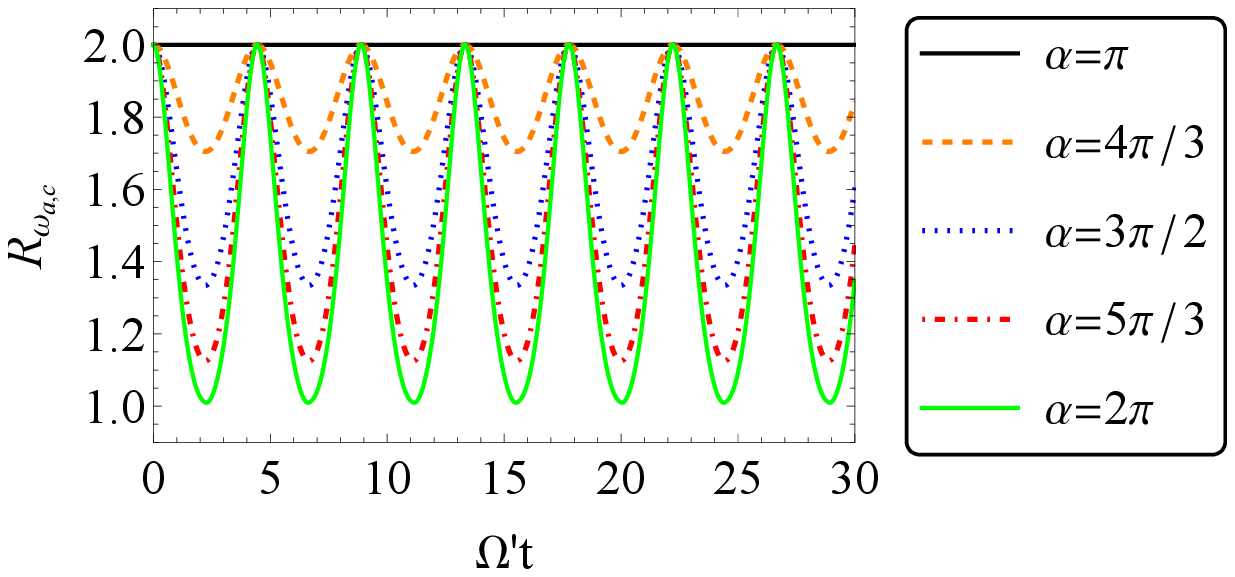}
	\caption{The variation  of
		performance ratio $R_{{\textrm{$\omega$}}_{\mathrm{a,c}}}$, calculated for simultaneous estimation of $ \omega_{a} $ and $ \omega_{c} $, as a function of the scaled time $ \Omega^{'}t $
		for ${\textrm{$\theta$}}$${=}{\textrm{$\pi$}}{/}{2}$
		and different values of
		${\textrm{$\alpha$}}$  in the equal Rabi frequencies regime.}.
	\label{Romegaac2}
\end{figure}
	
	In the  second scenario we intend to estimate the two upper  energy levels, i.e., $\{\omega_{a}, \omega_{c} \} $. Here it is seen  that the performance ratio $ R_{\omega_{a,c}} $ may be enhanced   by an increase in the Rabi frequency $ \Omega_{\mathrm{R2}} $ (see Fig. \ref{Romegaac1}). Hence,   when the driving field coupling the upper ground and excited states  is intensified considerably,   the simultaneous strategy becomes
	advantageous than the individual one. 
	
	\par
	Again we find that 
	  in the equal Rabi frequencies regime  if the coherent trapping occurs,
	  $ R_{\omega_{a,c}} $ is maximized, meaning complete superiority of the simultaneous estimation of the two upper energy levels compared to independent one (see Fig. \ref{Romegaac2}).

	\begin{figure}[ht]
		\includegraphics[width=8cm,height=5cm]{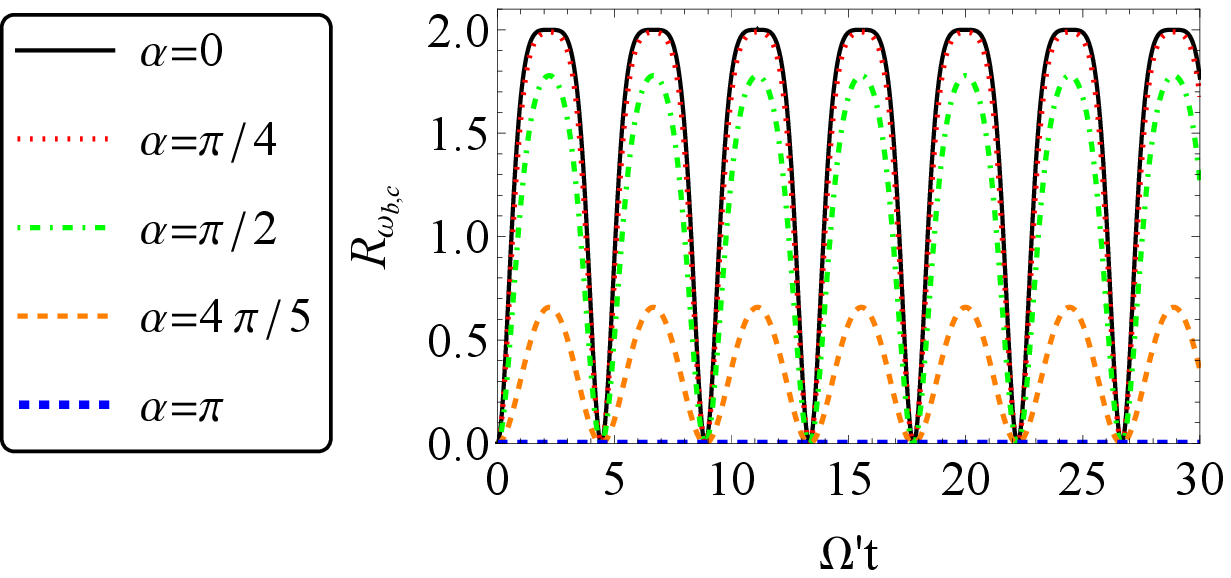}
		\hspace{10mm}
		\includegraphics[width=8cm,height=5cm]{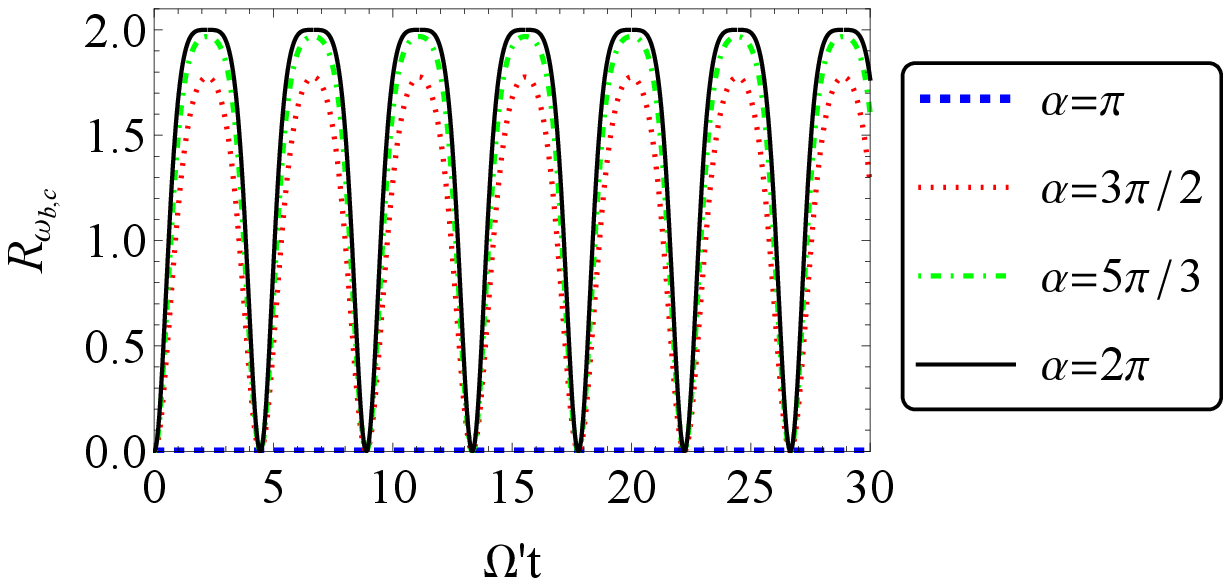}
		\caption{The variation  of
			performance ratio $R_{{\textrm{$\omega$}}_{\mathrm{b,c}}}$, calculated for simultaneous estimation of $ \omega_{b} $ and $ \omega_{c} $, as a function of the scaled time $ \Omega^{'}t $
			for ${\textrm{$\theta$}}$${=}{\textrm{$\pi$}}{/}{2}$
			and different values of
			${\textrm{$\alpha$}}$  in the equal Rabi frequencies regime.}.
		\label{Romegabc}
	\end{figure}

		\subsubsection{Simultaneous estimation of $ \omega_{b} $ and $ \omega_{c} $}
		
		Interestingly, in the two  previous subsections we saw that when the single upper level is simultaneously estimated with one of the lower levels, the best estimation is achieved provided that the coherent trapping occurs in the equal Rabi frequencies regime. Nevertheless, here we find  that the simultaneous estimation of the lower levels completely fails when the population is trapped in those  states, as shown in Fig. \ref{Romegabc} in which $ R_{\omega_{b,c}} $ is equal to zero in coherent trapping scenario. 
		\par 
		However, if we can control the Rabi phases $ (\phi_{1},\phi_{2}) $ as well as the initial phase $ \psi $ such that $ \alpha=0 $ is realized, the  corresponding QFIM becomes diagonal at times $ {\mathrm{\textrm{$\Omega$}}}^\prime t=\pi/\sqrt{2} $ for which the performance ratio $ R_{\omega_{b,c}} $ is maximized, indicating   completely supremacy of simultaneous estimation.

	\section{Conclusion and Discussion}\label{cunclusion}
 Applying multi-level systems, or qudits, instead of qubits to
implement quantum information tasks is a developing field promising advantages in some areas of quantum information.  In particular, recent theoretical
works suggest that quantum error correction with multi-level systems
has potential advantages over qubit-based schemes. With these considerations, determining the energy levels of  an atomic system is a crucial task to employ them in quantum information processing.
	\par
In this paper, considering a three-level atom interacting with laser fields, 	we examined the application of quantum metrology in  atomic spectroscopy focusing on detection of  spectra of atoms, ions, and sometimes molecular species based on measurements from both the electromagnetic and \textit{mass spectra}. Computing the quantum Fisher information matrrix (QFIM),  which is a key concept in quantum 
metrology, we investigated the effects of the initial preparation of the atom and the laser fields on energy levels estimation of the atom. It was shown that the influence of the laser phases on the quantum estimation can be controlled by the  phase  encoded into the initial state of the atom. Moreover, it was demonstrated that controlling the intensity of the laser fields, we can considerably enhance the estimation of the energy levels. In addition, the sensitive amplification enabled by coherent population trapping was discussed. Besides, we found that that there is a single measurement which is jointly optimal for extracting information on all
energy levels, intending to estimate them simultaneously,  ensuring the asymptotic saturability
of the Cramer-Rao bound.
\par
 We have also constructed an important relationship for the three level system  between the Hilbert–Schmidt 
speed (HSS), which is a special case of quantum statistical speed, and the QFI.
It was shown that these two quantities, computed with respect to the same energy levels, exhibit the same qualitative dynamical behaviors.
 This 
relationship originates from the fact that the QFI is itself a quantum statistical speed associated with the \textit{Bures distance}. 
 Moreover, in contrast to the typical computational complication of the QFI, especially 
for high dimensional systems, the HSS is easily computable and hence it can be instead proposed as a powerful tool 
to estimate energy levels in three-level systems. Our results pave the way to further 
studies on  applications of the HSS in quantum metrology as well as atomic spectroscopy.
	\subsection*{Acknowledgments}
	H. R. J. wishes to acknowledge the financial support of the MSRT of Iran and Jahrom
	University
	\appendix*
	
	\section{ Analytical expressions of Quantum Fisher information in single estimation scenario}
	
	The QFIs associated with energy levels are given by:
	\large
	\begin{equation}
		F_{{\textrm{$\omega$}}_{\mathrm{a\ }}}=t^2\Bigg(1-{(-1+\frac{{\mathrm{sin}[\frac{1}{2}t\sqrt{{\textrm{$\Omega$}}^2_{\mathrm{R1}}+{\textrm{$\Omega$}}^2_{\mathrm{R2}}}]}^2({\textrm{$\Omega$}}^2_{\mathrm{R1}}+2\:\mathrm{cos}\textrm{$\alpha$}\:\mathrm{sin}\textrm{$\theta$}\:{\textrm{$\Omega$}}_{\mathrm{R1}}{\textrm{$\Omega$}}_{\mathrm{R2}}+{\textrm{$\Omega$}}^2_{\mathrm{R2}}+\mathrm{cos}\textrm{$\theta$}({\textrm{$\Omega$}}^2_{\mathrm{R1}}-{\textrm{$\Omega$}}^2_{\mathrm{R2}}))}{{\textrm{$\Omega$}}^2_{\mathrm{R1}}+{\textrm{$\Omega$}}^2_{\mathrm{R2}}})}^2\Bigg),
	\end{equation}
	
	\begin{equation}
		\begin{split}
			F_{{\textrm{$\omega$}}_{\mathrm{b\ }}}&=t^2\Bigg(1-\frac{1}{{({\textrm{$\Omega$}}^2_{\mathrm{R1}}+{\textrm{$\Omega$}}^2_{\mathrm{R2}})}^4}{({({\textrm{$\Omega$}}^2_{\mathrm{R1}}+{\textrm{$\Omega$}}^2_{\mathrm{R2}})}^2}\\
			&+4\:\mathrm{cos}\textrm{$\alpha$}\:\mathrm{sin}\textrm{$\theta$}\:{\mathrm{sin}}^2[\frac{1}{4}t\sqrt{{\textrm{$\Omega$}}^2_{\mathrm{R1}}+{\textrm{$\Omega$}}^2_{\mathrm{R2}}}]{\textrm{$\Omega$}}_{\mathrm{R1}}{\textrm{$\Omega$}}_{\mathrm{R2}}(\mathrm{cos}[\frac{1}{2}t\sqrt{{\textrm{$\Omega$}}^2_{\mathrm{R1}}+{\textrm{$\Omega$}}^2_{\mathrm{R2}}}]{\textrm{$\Omega$}}^2_{\mathrm{R1}}+{\textrm{$\Omega$}}^2_{\mathrm{R2}})\\
			&-({\textrm{$\Omega$}}^2_{\mathrm{R1}}+{\textrm{$\Omega$}}^2_{\mathrm{R2}})({\mathrm{cos}}^2[\frac{1}{2}t\sqrt{{\textrm{$\Omega$}}^2_{\mathrm{R1}}+{\textrm{$\Omega$}}^2_{\mathrm{R2}}}]{\textrm{$\Omega$}}^2_{\mathrm{R1}}+{\textrm{$\Omega$}}^2_{\mathrm{R2}})-\mathrm{cos}\textrm{$\theta$}((-1+4\:\mathrm{cos}[\frac{1}{2}t\sqrt{{\textrm{$\Omega$}}^2_{\mathrm{R1}}+{\textrm{$\Omega$}}^2_{\mathrm{R2}}}]){\textrm{$\Omega$}}^2_{\mathrm{R1}}{\textrm{$\Omega$}}^2_{\mathrm{R2}}\\
			&+{\textrm{$\Omega$}}^4_{\mathrm{R2}}+{\mathrm{cos}}^2[\frac{1}{2}t\sqrt{{\mathrm{\textrm{$\Omega$}}^2_{\mathrm{R1}}}+{\textrm{$\Omega$}}^2_{\mathrm{R2}}}]({\textrm{$\Omega$}}^4_{\mathrm{R1}}-{\textrm{$\Omega$}}^2_{\mathrm{R1}}{\textrm{$\Omega$}}^2_{\mathrm{R2}})))^2\Bigg),
		\end{split}
	\end{equation}
	
	\begin{equation}
		\begin{split}
			F_{{\textrm{$\omega$}}_{\mathrm{c\ }}}&=t^2\Bigg(1-\frac{1}{{({\textrm{$\Omega$}}^2_{\mathrm{R1}}+{\textrm{$\Omega$}}^2_{\mathrm{R2}})}^4}{({({\textrm{$\Omega$}}^2_{\mathrm{R1}}+{\textrm{$\Omega$}}^2_{\mathrm{R2}})}^2+4\:\mathrm{cos}\textrm{$\alpha$}\:\mathrm{sin}\textrm{$\theta$}\:{\mathrm{sin}}^2[\frac{1}{4}t\sqrt{{\textrm{$\Omega$}}^2_{\mathrm{R1}}+{\textrm{$\Omega$}}^2_{\mathrm{R2}}}]{\textrm{$\Omega$}}_{\mathrm{R1}}{\textrm{$\Omega$}}_{\mathrm{R2}}({\textrm{$\Omega$}}^2_{\mathrm{R1}}}\\
			&+\mathrm{cos}[\frac{1}{2}t\sqrt{{\textrm{$\Omega$}}^2_{\mathrm{R1}}+{\textrm{$\Omega$}}^2_{\mathrm{R2}}}]{\textrm{$\Omega$}}^2_{\mathrm{R2}})+(-{\textrm{$\Omega$}}^2_{\mathrm{R1}}-{\textrm{$\Omega$}}^2_{\mathrm{R2}})({\textrm{$\Omega$}}^2_{\mathrm{R1}}+{\mathrm{cos}}^2[\frac{1}{2}t\sqrt{{\textrm{$\Omega$}}^2_{\mathrm{R1}}+{\textrm{$\Omega$}}^2_{\mathrm{R2}}}]{\textrm{$\Omega$}}^2_{\mathrm{R2}})\\
			&+\mathrm{cos}\textrm{$\theta$}(4\mathrm{cos}[\frac{1}{2}t\sqrt{{\textrm{$\Omega$}}^2_{\mathrm{R1}}+{\textrm{$\Omega$}}^2_{\mathrm{R2}}}]{\textrm{$\Omega$}}^2_{\mathrm{R1}}{\textrm{$\Omega$}}^2_{\mathrm{R2}}+({\textrm{$\Omega$}}^2_{\mathrm{R1}}-{\textrm{$\Omega$}}^2_{\mathrm{R2}})({\textrm{$\Omega$}}^2_{\mathrm{R1}}-{\mathrm{cos}}^2[\frac{1}{2}t\sqrt{{\textrm{$\Omega$}}^2_{\mathrm{R1}}+{\textrm{$\Omega$}}^2_{\mathrm{R2}}}]{\textrm{$\Omega$}}^2_{\mathrm{R2}})))^2\Bigg).
		\end{split}
	\end{equation}
		
	In  the  equal Rabi frequencies regime (${\mathrm{\textrm{$\Omega$}}}_{R1}=$ ${\mathrm{\textrm{$\Omega$}}}_{R2}$), the QFI expressions simplify as follows:
	
	\begin{equation}
		F_{{\textrm{$\omega$}}_{\mathrm{a\ }}}=-\frac{1}{2}t^2(1+\mathrm{cos}\textrm{$\alpha$}\:\mathrm{sin}\textrm{$\theta$}){\mathrm{sin}[\frac{t{\textrm{$\Omega$}^\prime}}{\sqrt{2}}]}^2\left(-3-\mathrm{cos}[\sqrt{2}t{\textrm{$\Omega$}^\prime}]+2\:\mathrm{cos}\textrm{$\alpha$}\:\mathrm{sin}\textrm{$\theta$}\:{\mathrm{sin}[\frac{t{\textrm{$\Omega$}}^\prime}{\sqrt{2}}]}^2\right),
	\end{equation}
	
	\begin{equation}
		\begin{split}
			F_{{\textrm{$\omega$}}_{\mathrm{b\ }}}&=\frac{1}{128}\:t^2\Bigg(81-\mathrm{cos}[2\textrm{$\theta$}](29+36\:\mathrm{cos}[\sqrt{2}\:t{\textrm{$\Omega$}^\prime}])-4\:\mathrm{cos}[2\sqrt{2}\:t{\textrm{$\Omega$}^\prime}]{(1+\mathrm{cos}\textrm{$\alpha$}\:\mathrm{sin}\textrm{$\theta$})}^2\\
			&-6\:\mathrm{sin}\textrm{$\theta$}(4\:\mathrm{cos}\textrm{$\alpha$}+\mathrm{cos}[2\textrm{$\alpha$}]\:\mathrm{sin}\textrm{$\theta$})+4\:\mathrm{cos}[\sqrt{2}\:t{\textrm{$\Omega$}^\prime}](-3+8\:\mathrm{cos}\textrm{$\alpha$}\:\mathrm{sin}\textrm{$\theta$}+2\:\mathrm{cos}[2\textrm{$\alpha$}]\:{\mathrm{sin}\textrm{$\theta$}}^2)\\
			&+128\:\mathrm{cos}\textrm{$\theta$}\:\mathrm{cos}[\frac{t{\textrm{$\Omega$}^\prime}}{\sqrt{2}}](1+\mathrm{cos}\textrm{$\alpha$}\:\mathrm{sin}\textrm{$\theta$}){\mathrm{sin}[\frac{t{\textrm{$\Omega$}^\prime}}{\sqrt{2}}]}^2\Bigg),
		\end{split}
	\end{equation}
	
	\begin{equation}
		\begin{split}
			F_{{\textrm{$\omega$}}_{\mathrm{c\ }}}&=\frac{1}{128}\:t^2\Bigg(81-\mathrm{cos}[2\textrm{$\theta$}](29+36\:\mathrm{cos}[\sqrt{2}\:t{\textrm{$\Omega$}^\prime}])-4\:\mathrm{cos}[2\sqrt{2}\:t{\textrm{$\Omega$}^\prime}]{(1+\mathrm{cos}\textrm{$\alpha$}\:\mathrm{sin}\textrm{$\theta$})}^2\\
			&-6\:\mathrm{sin}\textrm{$\theta$}(4\:\mathrm{cos}\textrm{$\alpha$}+\mathrm{cos}[2\textrm{$\alpha$}]\:\mathrm{sin}\textrm{$\theta$})+4\:\mathrm{cos}[\sqrt{2}\:t{\textrm{$\Omega$}^\prime}](-3+8\:\mathrm{cos}\textrm{$\alpha$}\:\mathrm{sin}\textrm{$\theta$}+2\:\mathrm{cos}[2\textrm{$\alpha$}]\:{\mathrm{sin}\textrm{$\theta$}}^2)\\
			&-128\:\mathrm{cos}\textrm{$\theta$}\:\mathrm{cos}[\frac{t{\textrm{$\Omega$}^\prime}}{\sqrt{2}}](1+\mathrm{cos}\textrm{$\alpha$}\:\mathrm{sin}\textrm{$\theta$}){\mathrm{sin}[\frac{t{\textrm{$\Omega$}^\prime}}{\sqrt{2}}]}^2\Bigg).
		\end{split}
	\end{equation}

	\bibliography{apssamp2}% Produces the bibliography via BibTeX.
	
\end{document}